\documentclass[pra,twocolumn,showpacs,superscriptaddress,amsmath, floatfix]{revtex4}

\usepackage{CJKutf8}
\usepackage[utf8]{inputenc}
\usepackage{bm}
\usepackage{amssymb}
\usepackage{graphicx}
\usepackage{dsfont}
\usepackage{epsfig}
\usepackage{color}
\usepackage{ulem}
\usepackage{soul,xcolor}
\usepackage{hyperref}
\usepackage{amsmath, amsthm, amssymb, amsfonts}
\usepackage{hep-math}
\usepackage{algorithmic} 
\usepackage{float}
\theoremstyle{definition}

\newcommand{\hide}[1]{}

\newcommand{\fig}[1]{Fig.\,\ref{#1}}

\graphicspath{{./figures/others}{./figures/discord}{./figures/entanglement}}
\begin{document}
\begin{CJK*}{UTF8}{min}
\title{ Inter-Particle Correlations in the Dissipative Phase Transition of a Collective Spin Model }
\author{Qingyang Seiyo Wang (王青陽）}
\email{qywangoh@gmail.com}
\affiliation{Department of Physics, Massachusetts Institute of Technology, Cambridge MA}

\author{Susanne F. Yelin}
\email{syelin@g.harvard.edu}
\affiliation{Department of Physics, Harvard University, Cambridge, MA}

\date{\today}
\begin{abstract}

In open quantum systems undergoing phase transitions, the intricate interplay between unitary and dissipative processes leaves many information-theoretic properties opaque. We are here interested in interparticle correlations within such systems, specifically examining quantum entanglement, quantum discord, and classical correlation within the steady state of a driven-dissipative collective spin model. This model is renowned for its transition from a high-purity to a low-purity state with decreasing dissipation. Our investigation, rooted in numerical analysis using PPT criteria, underscores that entanglement reaches its peak at the phase transition juncture. Intriguingly, within the mesoscopic scale near the transition point, entanglement endures across both phases, despite the open nature of the model. In stark contrast, quantum discord and its variations chart an alternate trajectory, ascending monotonically as the system progresses into the low-purity phase. Consequently, lowered dissipation amplifies quantum correlation, yet it engenders entanglement solely in proximity to the transition point.

%
\end{abstract}
\maketitle
\end{CJK*}

\section{Introduction and Motivation}

The study of fluctuations and phase transitions has been a central focus in physics and has entailed extensive theoretical and experimental research. By detecting and analyzing these fluctuations, alongside rigorous physical modeling, crucial advancements were achieved in observing, predicting, and comprehending phase transitions across a wide range of systems.

In both quantum and classical systems, these phase transitions and fluctuations are marked by divergent correlation lengths \cite{Sachdev2011,Vojta_2003} at the phase transition point, indicating slow decay of autocorrelation functions. In classical systems, this phenomenon is governed by probabilistic distributions of the states of individual particles, whereas in pure quantum systems, the source is quantum entanglement between the particles, as exemplified by the divergent entanglement entropy \cite{Osborne2002,Osterloh2002ScalingTransition}. A subsequent progression of such studies involves the examination of mixed systems where both types of correlation coexist. In equilibrium quantum systems at finite temperatures, their long-distance correlations are fundamentally classical \cite{Sachdev2011}, with quantum entanglement contributing minimally due to its typically short-range nature.

The interplay between the correlation of the classical and quantum nature becomes significantly more intricate in nonequilibrium dissipative quantum models. There, the system's dynamics is dictated by the Lindblad master equations. The balance between coherent driving and dissipation in these systems leads the system into a stationary state instead of a ground state, potentially giving rise to dissipative phase transitions with a distinct class of critical non-equilibrium phenomena in open quantum systems \cite{Minganti2021ContinuousBreaking}. Although there has been growing interest in analyzing quantum and classical fluctuations in DPTs \cite{Kessler2012, Boneberg2022QuantumModels, Casteels2017QuantumDimer}, comprehensive and general arguments for correlations around the phase transition still need to be explored. 

The difficulty of deciphering correlations in non-equilibrium systems can be largely attributed to the following reasons: (1) the probability distribution of the states, unlike those following Gibbs distribution in thermal equilibrium, is not predetermined and instead depends on the individual model being studied; (2) deriving a steady-state solution of a master equation proves to be challenging in general; and (3) quantifying correlation (including entanglement) in mixed-state systems poses computational difficulties, often resulting in NP-hard problems \cite{Huang2014ComputingNP-complete, Gharibian2008StrongProblem, Gurvits2004ClassicalEntanglement}.

Instead of attempting to offer a generalized framework for these complex issues, the present work focuses on a specific model as a case study on various correlation measures. An ideal model should (i) demonstrate phase transitions, (ii) be an open quantum mechanical model (contrasting a purely closed quantum system), and (iii) be sufficiently simple to allow elaborate calculations involved in correlation measures, essentially bypassing the challenges while preserving the distinctive properties inherent to nonequilibrium models. The driven-dissipative collective spin model \cite{Drummond1980, Walls1978Non-EquilibriumSystems} aligns with these criteria, making it a suitable choice for this study. The model mainly comprises two components; a unitary term that directs spins in one direction and a dissipative term that directs them in another direction. This combination of contradictory terms yields nontrivial open-system dynamics, leading to nonequilibrium stationary states that are qualitatively different from those of closed quantum systems. Furthermore, these conflicting terms are known to induce dissipative-phase transitions. The core dynamics can be examined using collective spin states (Dicke states), significantly minimizing the size of the Hilbert space, and the analytic form of the steady state is known. 

The manuscript is structured as follows. In \autoref{sec:model}, we detail the inherent properties and preexisting knowledge about the dissipative collective spin model. In \autoref{sec:entanglement}, we dive into the entanglement of the model. We report that the system is genuinely multiparticle entangled near the phase transition point. Finally, in \autoref{sec:discord}, we present the numerical results of quantum discord and its variants, revealing that quantum discord exhibits a markedly distinct behavior from entanglement.

\section{The Model: Driven-dissipative collective spin model}\label{sec:model}
 The model we consider in this manuscript is an open quantum system of collective spins in Dicke state with unitary evolution that energetically favors the x-direction, and dissipative evolution that favors the z-direction.  In Lindbladian master equation, this amounts to the Hamiltonian part $\hat{H} = \omega_R \hat{J}_x$ and the dissipative part with the collective jump operator $\hat{L} = \Gamma \hat{J}^-$, where $\hat{J}_x = \hat{J}^+ + \hat{J}^-$ and $\hat{J}^{ \pm}|J, m\rangle=[(J \mp m)(J \pm m+1)]^{1 / 2}|J, m \pm 1\rangle$ with $J_- = (J_+)^\dag$ are the collective spin operators. Explicitly, the master equation is 
\begin{equation}
    \frac{\partial \hat{\rho}}{\partial t}=-i[\omega_R \hat{J}_x, \hat{\rho}]+ \Gamma\left(\hat{J}^- \hat{\rho} J^+ -\frac{\hat{J}^{+} \hat{J}^{-} \hat{\rho}+\hat{\rho} \hat{J}^+ \hat{J}^-}{2}\right). \label{eq:dickemodel}
\end{equation}

Originally, the model was used to describe a collection of driven atoms in free space and their fluorescence dynamics \cite{Harrison1978,Drummond1980}. The model can be interpreted as 2-level atoms collectively interacting with a resonant semi-classical electromagnetic (E\&M) field characterized by a Rabi frequency $\omega_R$ with collective spontaneous emission whose collective spontaneous decay rate is denoted by $\Gamma$. Throughout the dynamics, atoms are assumed to be in the Dicke state, imposing complete permutation invariance. The dephasing of the atoms, which can arise from dipole-dipole interactions that break the symmetry, is assumed to be negligible. Applying the rotating wave approximation and the Born-Markov approximation \cite{Breuer2007TheSystems}, we arrive at \eqref{eq:dickemodel}.

Among the various proposed experimental realizations of the model, one particularly noteworthy set-up involves arranging qubits in a one-dimensional waveguide. This configuration facilitates controlled dephasing between the qubits, offering a compelling method for practical implementation at the mesoscopic scale.
\cite{Gonzalez-Tudela2013MesoscopicOptics}.

This model (\ref{eq:dickemodel}) is known to possess a unique steady state \cite{Harrison1978,Drummond1980}, which implies that there is always a unique solution to $\frac{\partial \rho}{\partial t} = 0$, irrespective of the parameter values $\omega$ and $\Gamma$. 
It is expressed as
\begin{align}
	\rho_{ss}&=\frac{1}{D}\sum_{l=0}^{2j} \sum_{l^{\prime}=0}^{2j} \left(\frac{J^{-}}{g}\right)^l\left(\frac{J^{+}}{g^*}\right)^{l^\prime} 
    \label{eq:exactsolution}
\end{align}
where $ g=i \omega_R/\Gamma $ and $D$ is the normalizing factor that ensures $\Tr(\rho) = 1$. 
In this paper, we focus exclusively on the steady state of the model. Consequently, from this point forward, the terms system and model used in this paper will refer to the steady-state solution of the model in \eqref{eq:exactsolution}, unless otherwise stated. We use $\hat{\rho}_{ss}$ for the steady state solution of the model and $\varrho$ for a generic density matrix. $\varrho$ would be used regularly to define quantities that involve a general density matrix not exclusive to $\hat{\rho}_{ss}$.

The model is primarily governed by two parameters: the ratio of the Rabi frequency $\omega_R$ to the collective decay rate $\Gamma$ and the number of particles $N$ ($=2J$). To simplify, we define a dimensionless parameter $\Omega = N\omega_R/\Gamma$ and mainly use $\Omega$ and $N$ to parameterize the steady state.

\begin{figure*}
  \centering
  \includegraphics[width=0.80\linewidth]{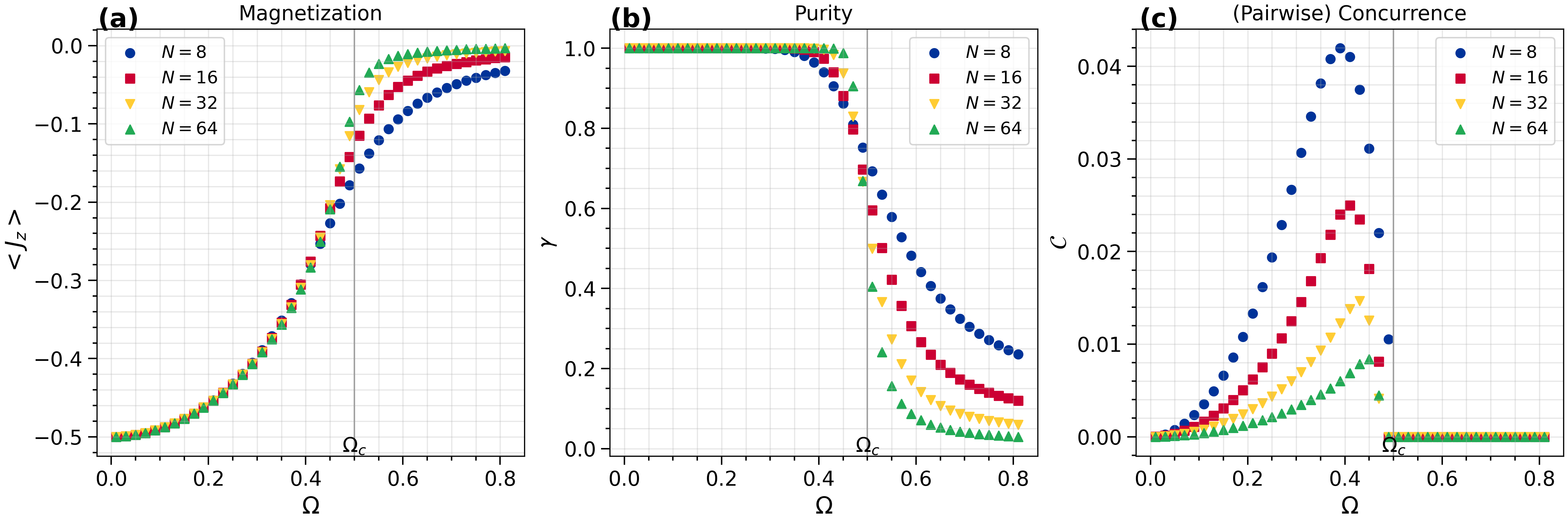}
  \caption{\label{fig:jz_purity_conc} (Color online) (a) Expected value of $J_z$, (b) Purity ($\gamma$), and (c) Pairwise concurrence ($C$) of the system are plotted as functions of $\Omega$ for varying particle numbers $N$ of the entire system. These plots replicate and validate the results previously reported. $\Omega (= N \omega_R / \Gamma) $ is the ratio of unitary term and dissipative term in \eqref{eq:dickemodel}. When the purity $\gamma = 1$, the system is in a pure state; when $\gamma < 1$, it is in a mixed state. A sharp decline in the purity of the system is observed for $\Omega > \Omega_c$, indicating a transition to a highly mixed state. Pairwise concurrence ($C$) is an entanglement monotone that quantifies the entanglement. Concurrence exhibits peaks near the transition point, while the height of the peak becomes smaller as the number of particles increases. The values of concurrence suddenly drop to zero in the impure phase, giving the impression that quantum entanglement is nonexistent in the mixed-state region. However, this is not the case, as we demonstrate in Section \ref{sec:entanglement}.  } 
\end{figure*}

When $N \to \infty$, the model is known to exhibit a dissipative phase transition (DPT) as $\Omega$ changes \cite{Harrison1978}.
A DPT is defined as a discontinuous change in $\rho_{ss}$ as a single parameter of the density matrix is continuously changed.  With $N$ fixed, the model is parameterized by $\Omega$, and signs of non-analyticity emerge when $\Omega = \Omega_C = 1/2$. For example, the first derivative of the expectation values of the angular momentum operators $<J_\alpha>$ ($\alpha = x, y, z$) shows a discontinuity at a critical point $\Omega_c$ (see \fig{fig:jz_purity_conc}). The scaling of the values are rather peculiar; $<J_z> \approx \sqrt{1-4\Omega^2}$ when $\Omega < \Omega_c$ but immediately becomes 0 for $\Omega>\Omega_c$. On the otherhand, $<J_y>$ is 0 when $\Omega < \Omega_c$ but scales as $<J_y> \approx 1/\Omega$ when $\Omega > \Omega_C$ \cite{Hannukainen2017DissipationBreaking}. 

Interestingly, the DPT shows a stark change in the purity of the system, $\gamma (= \Tr(\hat{\rho_{ss}}^2))$, as shown in \fig{fig:jz_purity_conc}. Contrary to intuition, the system remains pure when the dissipation term in \eqref{eq:dickemodel} is dominant ($\Omega < \Omega_C$). On the contrary, when the Hamiltonian term dominates ($\Omega > \Omega_C$), the system becomes highly mixed. Therefore, we call the phase for $\Omega < \Omega_C$, the pure phase, and $\Omega > \Omega_C$ impure phase.

Lastly, note that the phase transition in this model arises due to the simultaneous presence of a unitary component (represented by the Hamiltonian) and a dissipative component (represented by the jump operator) and does not align with the traditional Landau framework for continuous phase transitions, mainly because there is no obvious symmetry capable of spontaneous breaking \cite{Hannukainen2017DissipationBreaking}.  If the model had only two unitary terms, such as in the case where $H = \omega_R \hat{J_x} + \Gamma \hat{J_z}$, it would not show a phase transition. 

Up to this point, we have primarily focused on reviewing the known aspects of the driven-dissipative Dicke model. The following sections will present our findings derived from this understanding.

\section{Entanglement}
\label{sec:entanglement}

In closed quantum systems, peculiar behavior of entanglement signal phase transitions \cite{Osborne2002, Wu2004, Osterloh2002ScalingTransition}; however, in open systems, its role is more intricate due to environmental interactions that impact coherence and potential transitions.

The pioneering investigation of entanglement in this model used pairwise concurrence \cite{Schneider2002} (see \fig{fig:jz_purity_conc} (c) for the replication of the result.) This work revealed nonzero concurrence in the pure phase ($\Omega < \Omega_c$), with a peak near the critical point. However, the concurrence was observed to converge to zero as the number of particles increased. Subsequent studies similarly reported the presence of entanglement in the pure phase ($\Omega < \Omega_C)$, using spin squeezing \cite{Wolfe2014a}, as well as pairwise negativity and concurrence \cite{Hannukainen2017DissipationBreaking, Morrison2008}. At first glance, these results might suggest that atoms are quantum mechanically correlated (entangled) when the system is pure ($\Omega < \Omega_c$), with entanglement peaking at the transition point, but they completely lose quantum correlation as the probabilistic mixture becomes dominant ($\Omega > \Omega_c$), leading the system to become highly mixed. However, several points make the interpretation of entanglement from previous studies not entirely conclusive: (i) As the system scales to the thermodynamic limit, pairwise entanglement measures like concurrence trend towards zero, making the existence of entanglement less straightforward. Indeed, some suggest that due to the open nature of the model, entanglement could be completely destroyed at this limit \cite{Hannukainen2017DissipationBreaking}. (ii) Pairwise entanglement measures, computed on the reduced density matrix, may not fully represent the entanglement present in the entire system. (iii) Other measures employed in this model, such as entanglement entropy and spin squeezing, often fail to detect entanglement in mixed states. In light of these considerations, we are inspired to reexamine the model's entanglement.

The entanglement we discuss here is defined based on the separability criteria \cite{Aolita2015Open-systemReview}, which categorize any state that is not fully separable as entangled. For a mixed state of $N$ particles, represented by the density matrix $\varrho$, it is considered fully separable if and only if it can be expressed as a convex sum of pure $N$-product states: $\varrho = \sum _ { \mu } p _ { \mu } | \Psi _ { \mu _ { 1 } } \rangle \left\langle \Psi _ { \mu _ { 1 } } | \otimes \ldots \otimes | \Psi _ { \mu _ { N } } \right\rangle \left\langle \Psi _ { \mu _ { N } } |\right.$ where $| \Psi _ { \mu _ { i } } \rangle \in \mathcal { H } _ { i }$, $p_\mu >0$, and $\sum_\mu p_\mu = 1$. Here, $i$ labels each particle, and $\mu$ represents each pure product state. Any system that is not separable is thus termed as entangled. For systems with more than two particles, one could consider a system that is a convex sum of pure $k$-product states, given by $\rho = \sum _ { \mu } p _ { \mu } | \Psi _ { \mu _ { 1 } } ^ { \prime } \rangle \left\langle \Psi _ { \mu _ { 1 } } ^ { \prime } | \otimes \ldots \otimes | \Psi _ { \mu _ { k } } ^ { \prime } \right\rangle \left\langle \Psi _ { \mu _ { k } } ^ { \prime } |\right.$.
This is referred to as a $k$-separable state. Evidently, a fully separable state is an $N$-separable state. Any state that is not 2-separable is called genuinely multiparticle entangled (GME).

In order to develop a deeper understanding of the existence and structure of entanglement in this model based on the above definitions, we focus primarily on examining the entanglement using (a) the positive partial transpose (PPT) criterion \cite{Peres1996, Horodecki1996} and (b) the PPT mixture criterion \cite{Jungnitsch2010}.

\subsection{PPT criterion and negativity}
For a matrix $\varrho_{j_A, j_B, j_A^\prime j_B^\prime}$, given subsystems A and B, the partial transpose matrix $\varrho^{T_B}_{j_A, j_B, j_A^\prime j_B^\prime}$ is defined as

\begin{align}
\varrho_{j_{A} j_{B} j_{A}^\prime j_{B}^\prime}^{T_{B}}
&\equiv\left\langle j_{A}\left|\left\langle j_{B}\left|\varrho^{T_{B}}\right| j_{A}^{\prime}\right\rangle\right| j_{B}^{\prime}\right\rangle \\
&=\left\langle j_{A}\left|\left\langle j_{B}^{\prime}|\varrho| j_{A}^{\prime}\right\rangle\right| j_{B}\right\rangle \\
&= \varrho_{j_{A} j_{B}^{\prime} j_{A}^{\prime} j_{B}}
\end{align}
PPT states are $\varrho$ such that the eignevalues of $\varrho^{T_{B}}$ are all positive. 
 The Peres-Horodecki criterion asserts that any separable state is PPT. The inverse is not always true. Consequently, taking the contrapositive, if $\varrho$ is not PPT, it can be concluded that $\varrho$ cannot be separated as $\varrho = \varrho_A \otimes \varrho_B$ and by definition there is an entanglement between A and B. The commonly used measure to test the Peres-Horodecki criterion is negativity  ($\mathcal{N}$)  which is an entangle monotone \cite{Eisert2006} \cite{Vidal2001}.
  It is defined as the absolute value of the sum of negative eigenvalues of $\varrho^{T_A}$.
\begin{equation}
    \mathcal{N}(\rho)=\left|\sum_{\lambda_i<0} \lambda_i\right|
\end{equation}

$\mathcal{N} > 0$ is a sufficient condition for the state $\rho$ to be entangled. However, $\mathcal{N} = 0$ does not necessarily imply that the system is separable. It is well known that there exist states which are positive partial transpose (PPT) yet still entangled, known as PPT entangled states (PPTES) \cite{Horodecki1998Mixed-StateNature}. The search for PPTES is a challenging problem \cite{Augusiak2012EntangledTranspositions, Park2023ASymmetry} that lies beyond the scope of this work. Consequently, the subsequent discussion will be limited to the entangled states that can be detected by the PPT criterion, specifically, the negative partial transpose states (NPT).

To fully implement the PPT criterion on the entire density matrix using negativity, one needs to calculate the negativity for all bipartitions of the system. In a general quantum system consisting of $N$ atoms (or qubits), there are $2^{N }-1$ such partitions. Each bipartition requires the calculation of eigenvalues for a density matrix of size ($2^N \times 2^N$). This quickly leads to intractable computations. However, our model requires only $\lfloor N/2 \rfloor$ bipartitions ($\lfloor \quad \rfloor$ denotes the floor function, which returns the maximum integer). This simplification results from the permutation symmetry of the atoms, indicating that the particle number in each subsystem is enough to characterize the partition. Additionally, when using the Dicke basis, the size of the density matrix is reduced to $(N+1) \times (N+1)$ by defining quantum states in $\mathbb{C}_{N+1}$.  The partial transpose disrupts the permutation symmetry of the whole system, but symmetry is preserved within each Hilbert space supporting the state of the individual subsystems. Therefore, for each bipartition, where $N_A$ and $N_B$ represent the particle numbers of subsystems A and B respectively, we reshape the density into a size of $(N_A+1)(N_B+1) \times (N_A+1)(N_B+1)$ by treating the quantum state as bipartite symmetric state in the tensor product space $\mathbb{C}^{N_A+1} \otimes \mathbb{C}^{N_B+1}$. The partial tranpose of the density matrix, in this configuration, still acts on the same Hilbert space as the original density matrix. 

To fully implement the PPT criterion on the entire density matrix using negativity, one must calculate the negativity for all bipartitions of the system. For a generic quantum system comprising \(N\) atoms (or qubits), there exist \(2^{N}-1\) such partitions. Each bipartition requires the calculation of eigenvalues for a density matrix of dimensions \(2^N \times 2^N\). This quickly leads to intractable computations. However, in our model, only \(\lfloor N/2 \rfloor\) bipartitions are needed, where \(\lfloor \cdot \rfloor\) denotes the floor function that returns the maximum integer less than or equal to a given number. This simplification results from the permutation symmetry of the atoms, indicating that the particle number in each subsystem is enough to characterize the partition. Furthermore, using Dicke basis, we reduce the size of the density matrix to \((N+1) \times (N+1)\), defining quantum states in \(\mathbb{C}^{N+1}\). While the partial transpose disrupts the permutation symmetry of the entire system, symmetry remains intact within the Hilbert space on which individual subsystem states are defined. Consequently, for each bipartition, where \(N_A\) and \(N_B\) denote the particle numbers for subsystems A and B, respectively, we reshape the density matrix into dimensions \((N_A+1)(N_B+1) \times (N_A+1)(N_B+1)\). Here, we interpret the quantum state as a bipartite symmetric state in the tensor product space \(\mathbb{C}^{N_A+1} \otimes \mathbb{C}^{N_B+1}\). In this representation, the partial transpose of the density matrix remains defined on the same Hilbert space as its original counterpart.
With this approach, the overall computational efficiency of the PPT test in our model scales as $O(N E(N^2))$, contrasting with the $O(2^N E(2^N))$ scaling in the general case, where $E(d)$ denotes the complexity to obtain the eigenvalues for matrix of size $d \times d$. This improved efficiency enables us to detect entanglement within the full density matrix on a mesoscopic scale, and observe its scaling with the particle number $N$, thereby inferring the thermodynamic limit.

\begin{figure}[h!]
  \centering
  \includegraphics[width=0.9\columnwidth]{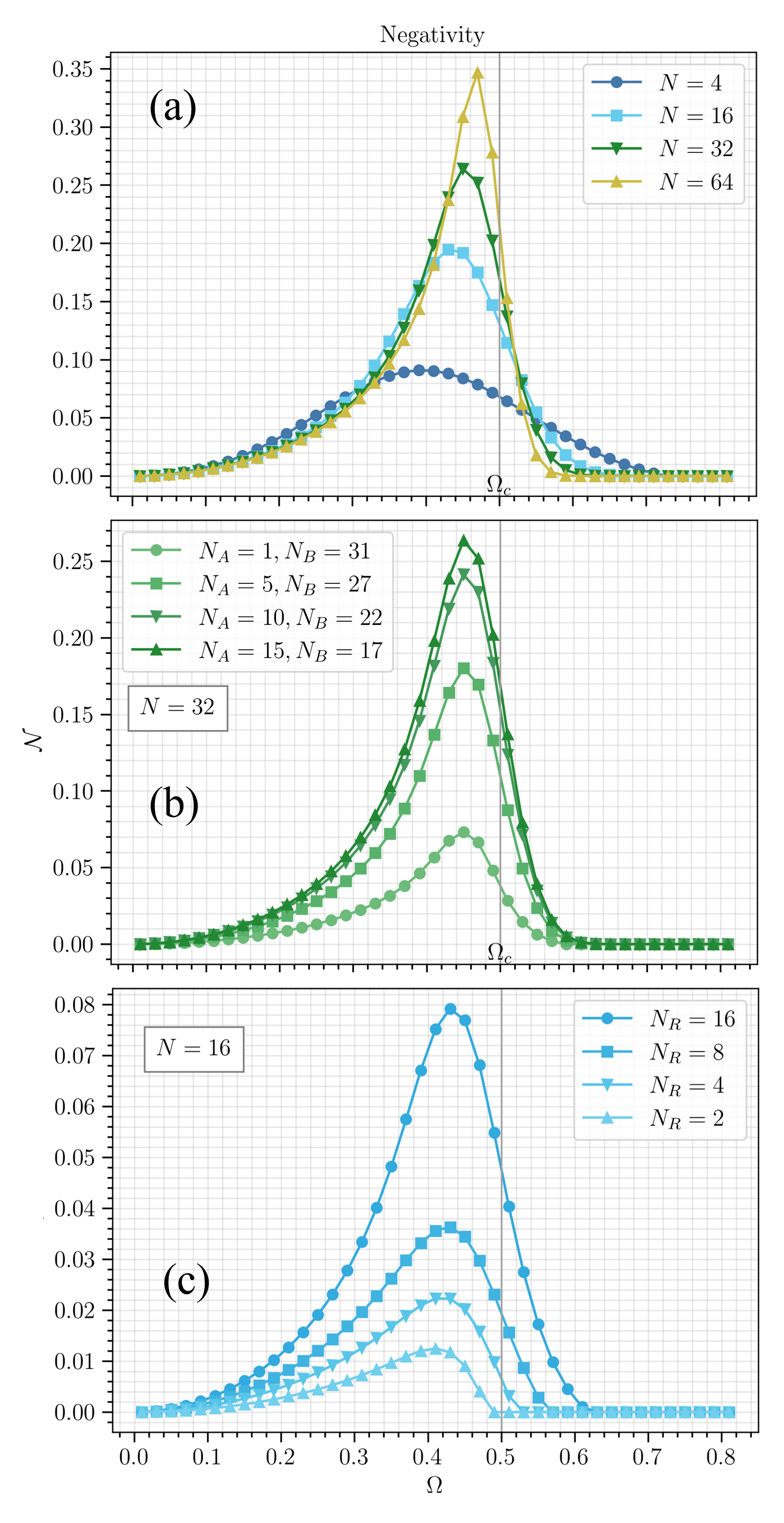}
  \caption{\label{fig:negativity} (Color online) Negativity of the system as a function of $\Omega$. (a) $N = N_R = 2, 16, 32, 64$. $N_A = N_B = N/2$,  (b) $N= N_R = 32$, $N_A = 1, 5, 10, 15$   (c) $N=16$, $N_R = 16, 8, 4, 2$, $N_A = 1$. (a) shows the scaling of negativity versus the number of particles. (b) shows the effect of different bipartition of subsystems when computing negativity. (c) shows the effect of tracing operation on the density matrix on negativity. From (b) We observe that negativity is overall the same upto the magnitude versus each bipartition of the system. From (c) we observe that tracing the density matrix significantly affect the entanglement in the region of $\Omega > \Omega_c$ only.   } 
\end{figure}

In Figure \fig{fig:negativity}, we have plotted the negativity under various conditions. Here, $N$ represents the total number of particles in the system, $N_R$ denotes the number of particles in the reduced density matrix, and $N_A$ is the subsystem on which the partial transpose operation is executed. $N_B$ represents the remaining particles. $N_A + N_B = N_R$, and $N_R = N$ if no trace operation is performed on the density matrix. We observe that, although the overall magnitude of negativity changes, the shape of the curve is largely unaffected by different partitions of the subsystems (\fig{fig:negativity} (b)). 

 Notably, the peak of negativity increases with the size of the system. This is exemplified when $N_A = N/2$ and $N_B = N/2$ (\fig{fig:negativity} (a)). Hence, we infer that entanglement exists and peaks at the transition point even in the thermodynamic limit, which was not supported by the decaying peak observed in the pairwise entanglement measures (see \fig{fig:jz_purity_conc}). 

Interestingly, on a mesoscopic scale (i.e., when $N < 100$), we observe that the system retains its entanglement for $\Omega_C > \Omega$ when $\Omega$ is close to $\Omega_c$, despite the system transitioning to a highly mixed state. This observation was not reported in previous studies that employed spin squeezing and pairwise entanglement measures. However, the detection of entanglement in this region is significantly hampered by trace operations. This phenomenon is evident in Figure \fig{fig:negativity} (c); as the number of particles that are traced out increases, negativity abruptly trends towards 0 when $\Omega_C > \Omega$. This behavior indicates a sudden shift in the entanglement class at the transition point. 

\subsection{PPT mixture witness}
While the results from negativity can detect entanglement, generally, they do not reveal the nature of how the system is entangled. Specifically, negativity does not provide the separability class of the system. However, in our model, any entanglement identified via the PPT criterion is genuinely multiparticle entangled (GME).

This can be intuitively explained as follows: Consider a pure Dicke (symmetric) state that is separable with respect to a bipartition of subsystems $A$ and $B$. Given an arbitrary pair of particles, denoted by $l$ and $m$, we can always ensure $l \in A$ and $m \in B$ by swapping the particles. This indicates that any two particles can belong to two distinct subsystems separable from each other. Consequently, every particle is separable from all others, implying full separability (refer to \cite{Ichikawa2008ExchangeEntanglement} for a formal discussion).
For a mixed state, the separable density matrix of a Dicke state is a convex sum of separable pure Dicke states, each of which is fully separable on its own. Hence, the density matrix must be fully separable if it is separable. Therefore, the density matrix of Dicke states is either not separable (GME) or fully separable.

For the consistency of our numerical results, we explicitly confirm the GME using the PPT mixture criterion \cite{Jungnitsch2010}, an extension of the PPT criterion, which is used to detect genuine multiparticle entanglement (GME) of a system. The underlying concept is to explore whether a given density matrix can be expressed as a (probabilistic) mixture of PPT states. Such states can be formally written as
\begin{equation}
    \varrho = \sum_i p_i \varrho^{PPT}_i
    \label{eq:pptmixture}
\end{equation}
where index $i$ denotes each bipartition of the system and $p_i$ are probabilistic weights. The PPT criterion we have studied in the previous subsection is essentially the case when $p_i = 1$. However, the PPT mixture criterion extends it to general $p_i$. Any $\varrho$ that is a mixture of PPT is 2-separable. On the contrary, if such a mixture is not found, the system is not 2-separable, and therefore the system is GME. 

We adopted the approach presented in \cite{Jungnitsch2010} where we explore the fully decomposable witness $W$ that is positive for all PPT mixture states. Such a witness provides a necessary condition for a 2-separable state or, taking the contrapositive, a sufficient condition for GME: 
\begin{equation}
    \operatorname{Tr}(W \varrho) < 0 \Rightarrow \varrho \ \text{is GME}
\end{equation}
The witness can be explored as follows:
\begin{align*}
& {\text{minimize}} & &  \text{Tr}(\rho W) \\
& \text{subject to} & & \text{Tr}(W) = 1 \\
& & & W = P_S + Q_S^{T_S}, \quad \forall S \\
& & & Q_S \geq 0, \quad P_S \geq 0 
\end{align*}
with $S$ denoting bipartitions of the subsystem. This optimization problem is situated within the realm of semidefinite programming and can be efficiently resolved. We utilized the PICOS and CVXPY for this task. The numerical results were cross-verified using the PPTmixer package introduced in \cite{Jungnitsch2010}, coupled with the YALMIP package and SDPT3 as the solver. Using the witness, one can construct an entanglement monotone known as genuine multiparticle negativity (GMN): $N_g = - \Tr(\varrho W)$ if $ \Tr(\varrho W) < 0$ and 0 otherwise \cite{Jungnitsch2010}. 

\begin{figure}[h]
\includegraphics[width=0.9\columnwidth]{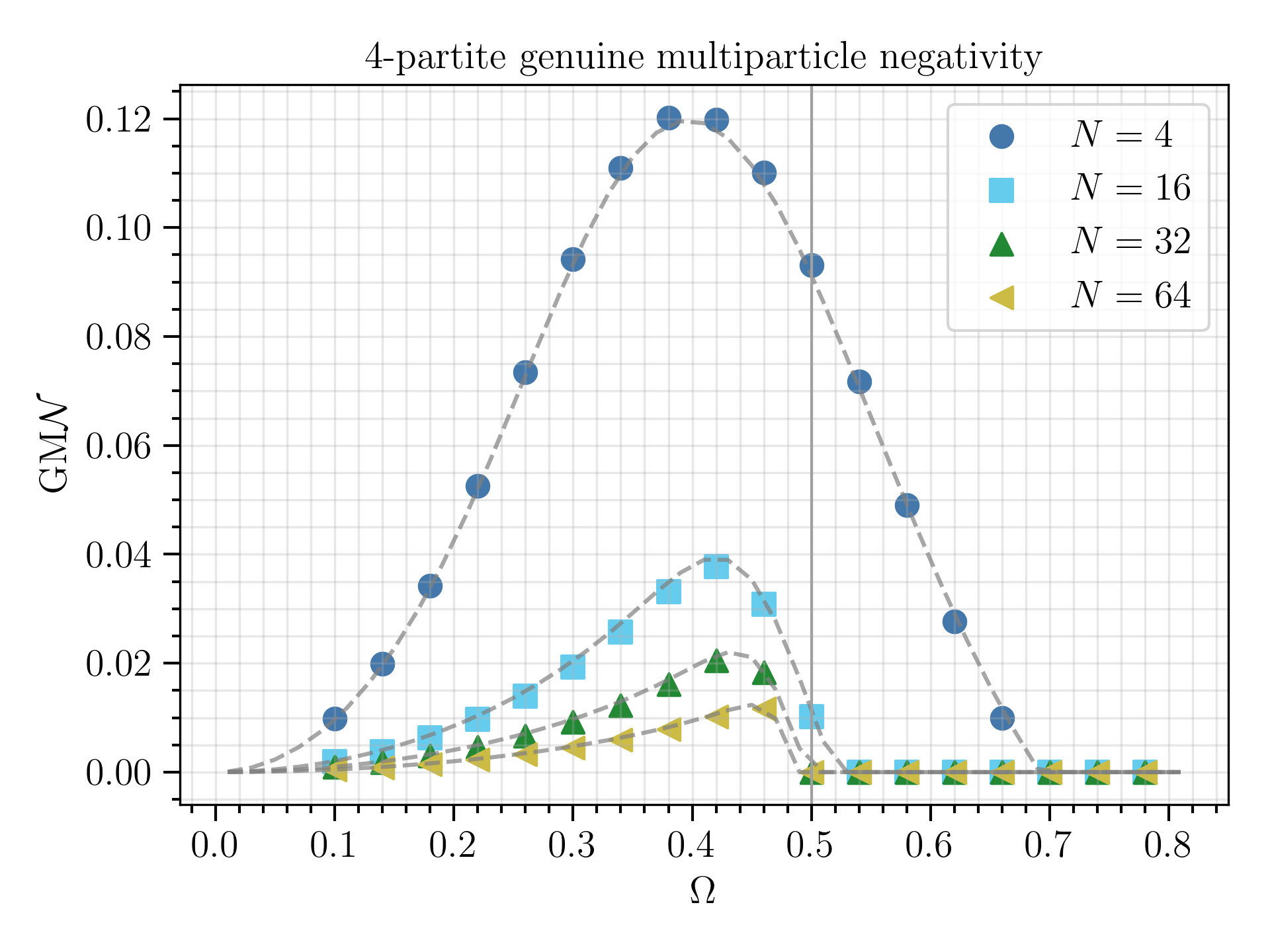}
\caption{\label{fig:gme_negativity}(Color online) 
Genuine multiparticle negativity of the system plotted against $\Omega$ for the reduced density matrix $N_R = 4$ derived from an original system with a particle number $N=4, 16, 32, 64$. The dashed line is the rescaled bipartition negativity computed for the same system size. }
\end{figure}

The plot of GMN for our model is shown in \fig{fig:gme_negativity}. in general, the entangled state detected by bipartite negativity does not correspond to that of GMN (Imagine a 2-separable entangled state $\varrho = p_1 \varrho_{A|BC} + p_2\varrho_{B|CA} + p_3 \varrho_{C|AB}$. This state may be detected as entangled by bipartite negativity, but is not detected by GMN). However, in our case, the overall behavior of GMN and bipartite negativity are almost identical, confirming that the entanglement detected by bipartite negativity is GME.

\subsection{Entanglement at two limiting cases}
Since the PPT-based criteria typically present a sufficient condition for the existence of entanglement, they do not produce any information on separability and entanglement when they fail to detect entanglement. Consequently, it remains inconclusive whether the system is entangled when $\Omega$ is far from the critical point $\Omega_C$  when both the negativity and PPT mixture witness are 0. Here, we demonstrate that the system is indeed separable and therefore not entangled in the two extreme cases: $\Omega \to 0$ and $\Omega \to \infty$. 

As $\Omega \to 0$ (dissipation dominant), the solution to Eq.~\eqref{eq:dickemodel} converges to \begin{equation} \lim_{\Omega \to 0} \rho_{ex}(\Omega) = \ket{J,-J}\bra{J,-J}. \end{equation} This state is clearly separable and is not entangled because $\ket{J,-J} = \ket{g_1} \otimes \ket{g_2} \cdots \otimes \ket{g_{2J}}$.

As $\Omega$ approaches infinity, the solution to Eq.~\eqref{eq:dickemodel} converges to \begin{equation} \lim_{\Omega \to \infty} \rho_{ex}(\Omega) = \frac{1}{N}\sum_{m = -J}^{J} \ket{J, m} \bra{J, m}. \label{eq:infsol} \end{equation} This state belongs to the class of diagonal symmetric states, which are composed of probabilistic mixtures of Dicke states. It is known that such states are separable if and only if they are PPT (Positive Partial Transpose) \cite{Yu2016}. Consistent with this, our numerical results demonstrate that the negativity is zero for all subsystem partitions as $\Omega$ approaches infinity, suggesting that the state is PPT and, therefore, separable. We also offer an analytical proof, affirming that $\rho_{ss}(\Omega \to \infty)$ is separable, regardless of the number of particles (refer to Appendix \ref{app:proof_max_mix} for details).

In summary, the steady state of the dissipative Dicke model is characterized by multipartite entanglement in the vicinity of the critical point $\Omega_C$. On the mesoscopic scale, entanglement is present for both $\Omega > \Omega_C$ and $\Omega < \Omega_C$, peaking at the transition point, as evidenced by the negative partial transpose (NPT) of the density matrix. In the thermodynamic limit, entanglement for $\Omega > \Omega_C$, which can be detected by PPT criteria, ceases to exist. Furthermore, when $\Omega$ is significantly divergent from $\Omega_C$, the density matrix is separable and, therefore, not entangled.

\section{Quantum Discord of the model} \label{sec:discord}
Although quantum entanglement has played a central role in contemporary quantum physics \cite{Mattle1996DenseCommunication, Srensen2001EntanglementSqueezing, Northup2014QuantumPhotons}, it is not the only form of quantum correlation. The entanglement formalism discussed in the previous sections focused only on the separability of the physical system and did not address the concept of correlation. Quantum discord represents an alternative approach to quantum correlation \cite{Zurek2000}.

Measurement-based quantum discord (QD) is fundamentally the difference between two classically equivalent definitions of mutual information\cite{Vidal2001, Henderson2001ClassicalCorrelations}. In probability theory, given random variables $A$ and $B$, the mutual information between $A$ and $B$ ($I(A: B)$) is:
\begin{align}
I(A: B) &= H(A)+H(B)-H(A, B) \label{eq:total_correlation} \\
  &= H(B) - H(B\mid A) \label{eq:cond_correlation}
\end{align}
where $H(A), H(B), H(A,B)$ and $H(B\mid A)$ are the Shannon entropy of marginal distributions over $A$, $B$, the joint distribution over $A\times B$ and conditional entropy, respectively.  To extend this into the quantum domain, we replace the Shannon entropy with the von Neumann entropy ($S$), and the probability distributions with the quantum density matrix ($\varrho$). The quantum version of \eqref{eq:total_correlation}, yields a total correlation.
\begin{equation}
\mathcal{I}_{AB}=S(\varrho_A)+S(\varrho_B)-S(\varrho_{AB}) \label{eq:discord}
\end{equation}

However, equation \eqref{eq:cond_correlation} is ill defined in quantum mechanics because making a measurement on a system inherently perturbs it (a phenomenon called the back action), causing the quantum version of $H(B\mid A)$ to be influenced by the choice of such measurements. Considering all possible sets of measurements, it is intuitive to assume that the state that is least perturbed by all such measurements would represent the "most classical" state. Consequently, the following quantity, an extension of equation \eqref{eq:cond_correlation}, should be considered as a measure of classical correlation.
\begin{equation}
\mathcal{J}_{AB}(\varrho)=S(\varrho_B)-S_{B/A}
\label{eq:condentropy}
\end{equation}
where
\begin{equation}
S_{B \mid A}=\min_{\Pi^A_k \in \mathcal{M}^A} \sum_k p_k S(\rho_{B/k})
\end{equation}
is the minimization of the conditional entropy with respect to all the measurement set on subsystem A. Here, $\pi_k^A$ is the measurement operator performed on the subsystem $A$ with a possible output $k$. $\mathcal{M}^A$ is the set of all such possible operators, including all possible $k$. $p_k=\text{tr}{AB}[I^B_m\otimes\Pi^A_k \rho{AB} , I^B_m \otimes (\Pi^A_k )^{\dag}]$ is the probability of the outcome of the measurement being $k$, and
$\rho_{B/k}=\text{tr}{A}[I^B_m\otimes\Pi^A_k, \rho{AB} , I^B_m \otimes (\Pi^A_k )^{\dag}]/p_k$ is the density matrix of the subsystem $B$ after the measurement. The difference of $\mathcal{I}_{AB}$ and $\mathcal{J}_{AB}$ is believed to produce a quantum correlation, and this is the quantum discord ($D^{\rightarrow}(\rho_{AB}$)).
\begin{align}
\mathcal{D}^{\rightarrow}(\varrho_{AB}) &\equiv \mathcal{I}_{AB}-\mathcal{J}_{AB} \nonumber \\
&= S(\varrho_A)+S_{B/A}-S(\varrho_{AB}) \label{eq:cond_entropy}
\end{align}

Quantum discord is non-zero for entangled states, yet it can exhibit nonzero values even for separable states. Therefore, it captures quantum correlations beyond entanglement, expanding our understanding and potential applications of nonclassical correlations. It has shown robustness against noise and decoherence, which makes it valuable for quantum information processing tasks where entanglement is fragile \cite{Smolin2012EfficientNoise, Werlang2009RobustnessDeath}. Furthermore, it plays a role in quantum computation and other information tasks even without entanglement \cite{Datta2008QuantumQubit}, and has connections to various aspects of phase transitions \cite{Mazzola2010SuddenDecoherence, Fan2013QuantumChain, Huang2014ScalingModels, Werlang2010QuantumTransitions}.

In our study, we computed 2-qubit quantum discord of the reduced density matrix, 2-partite QD of the full density matrix, and the global quantum discord of the steady state of the dissipative-Dicke model to reveal the classical and quantum correlation over the phase transition beyond entanglement.

\subsection{2-qubit Quantum Discord}
The calculation of quantum discord can be quite complex, particularly for high-dimensional systems, and its operational interpretation remains a topic of active investigation. However, significant progress has been made in determining the quantum discord for two-qubit systems \cite{Huang2013QuantumError}. For two-qubit systems, it has been demonstrated that considering sets of von Neumann measurements (as opposed to general POVM measurements) on one of the qubits suffices for the optimization process involved in \eqref{eq:cond_entropy} \cite{}. Typically, such von Neumann measurements are defined by the rotation of the eigenstates of the Pauli matrix $\sigma_z$; in our case, these are $\ket{e}$ (the excited state) and $\ket{g}$ (the ground state) of the atom. Explicitly, the measurement sets are described by $\ket{+}\bra{+}$ and $\ket{-}\bra{-}$, where 
\begin{align}
|+\rangle & = \mathrm{cos} \left(\frac{\theta}{2}\right)|e\rangle +\mathrm{e}^{i \phi} \sin \left(\frac{\theta}{2}\right)|g\rangle, \label{eq:plus}\\
|-\rangle & =-\mathrm{e}^{-i \phi} \sin \left(\frac{\theta}{2}\right)|e\rangle+\mathrm{cos} \left(\frac{\theta}{2}\right)|g\rangle. \label{eq:minus}
\end{align}

$\theta$ and $\phi$ are the parameters governing the rotation. $\mathrm{cos}$

\begin{figure}[h]
\centering
\includegraphics[width=0.9\columnwidth]{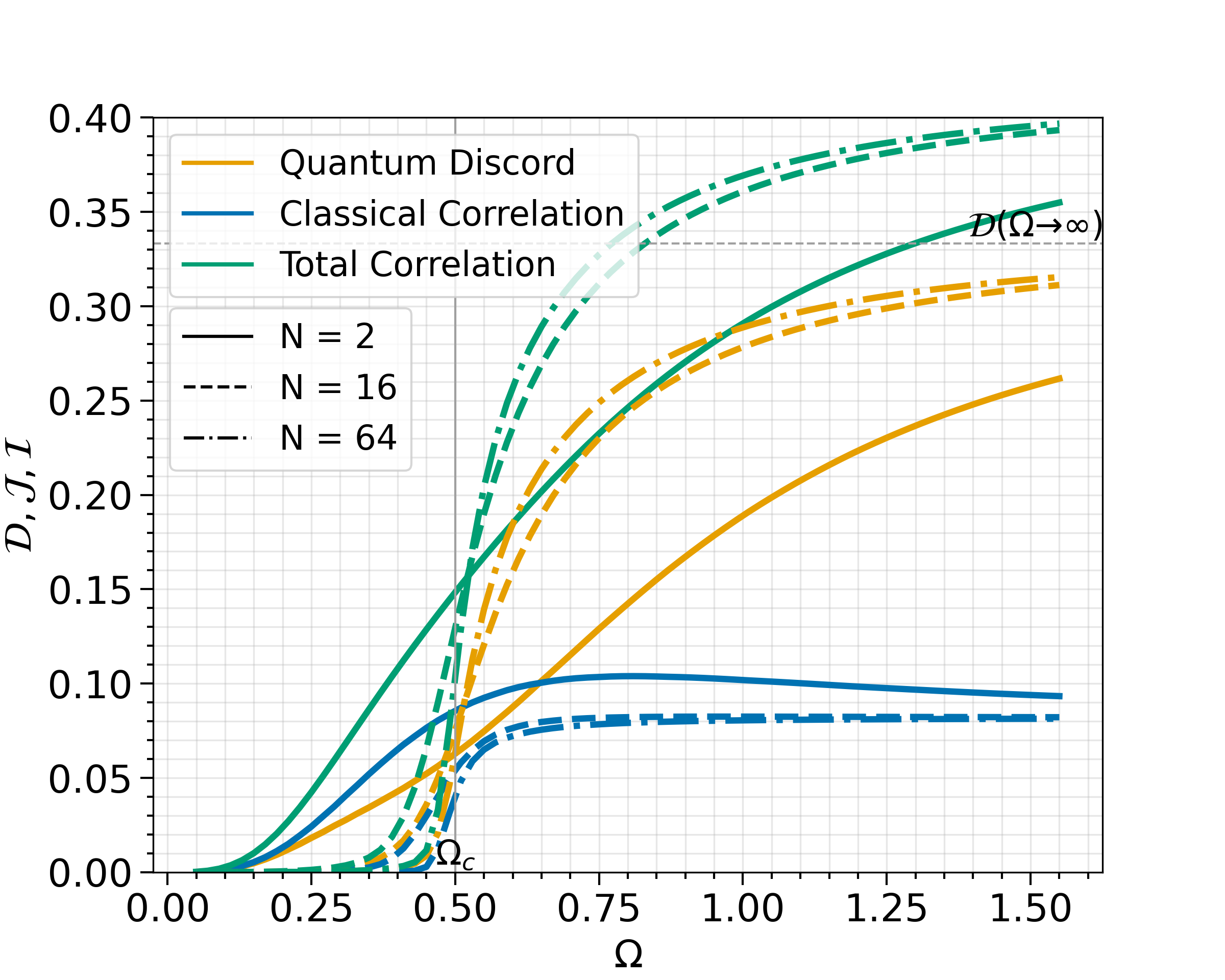}
\caption{\label{fig:dq_qubit_sum} (Color online) 2-qubit quantum discord (blue circle), classical correlation (light blue square), and total correlation (green triangle) plotted for different particle number $N$. (a) $N=2$, (b) $N=8$, (c) $N=16$.  For the computation, the full density matrix is traced out to obtain the 2-qubit density matrix. At $\Omega \to \infty$, QD converges to analytical value $1/3$.  }
\end{figure}

In this study, we first acquire the two-qubit density matrix by tracing out $N-2$ particles for each $\Omega$. Then, we numerically minimize \eqref{eq:cond_entropy} to obtain the two-qubit quantum discord. The results are shown in \fig{fig:dq_qubit_sum}.

We observe that, in contrast to the entanglement results, the quantum discord (and total correlation) monotonically increases and approaches the analytical value of QD at $\Omega \to \infty$. This analytical value, $\mathcal{D} = 1/3$ in our case, is derived from the general result of quantum discord for X states \cite{Fanchini2009}, as the density matrix of our system converges to the state when $\Omega \to \infty$.

We note that classical correlation, the difference between total correlation and quantum discord, does not monotonically increases, but rather peaks at the vicinity of the transition point ($\Omega_C$) in the impure phase. 

\subsection{bipartite quantum discord beyond 2-qubit systems}

As noted in Section \ref{sec:entanglement}, tracing out an extensive number of particles can result in significant information loss about the overall system. This issue became evident during our exploration of negativity, where the entanglement detection was substantially constrained due to the tracing operation. Bearing this experience in mind for the present section, we approached the problem cautiously to avoid similar information loss. We computed the upper limit of bipartite quantum discord for an increased particle count, all the while making an effort to minimize the number of atoms being traced out.

The system we studied is spanned by states with an effective angular momentum quantum number $J$, which let us consider it as bipartite. In this scenario, subsystems A and B have quantum numbers $j_a$ and $j_b$ respectively, where $j_a + j_b = J$. This connection arises because the angular momentum quantum number relates to the number of particles $N$ in the system through the relationship $J = N/2$, and the total number of particles remains conserved in the model.

Given $J = j_a + j_b$, a state $\ket{J,M}$ with overall angular momentum $J$ and its projection $M$ expands as:
\begin{equation}
\ket{J,M} =\sum_{a,b} C(J;j_a,j_b,M;a,b)\ket{j_a,a} \ket{j_b,b}
\end{equation}
where $C(J;j_a,j_b,M;a,b)$ are the Clebsch-Gordan coefficients. Using this expansion, we can break down the density matrix represented in the total $J$ basis as follows:

\begin{align}
\rho &= \sum_{M,M^\prime} \rho_{M,M^\prime}\ket{J,M}\bra{J,M^\prime} \\
&= \sum_{a,a^\prime,b,b^\prime} \left( \sum_{M,M^\prime} C(J;j_a,j_b,M;a,b) C(J;j_a,j_b,M^\prime;a^\prime,b^\prime) \right) \\ 
& \times \ket{j_a,a} \bra{j_a^\prime, a^\prime} \otimes \ket{j_b, b} \bra{j_b,b^\prime}
\end{align}

Ideally, we could optimize the conditional entropy by examining all possible sets of POVM measurements performed on the Hilbert space. However, this process becomes significantly complex for general bipartite systems with spins greater than $1/2$ \cite{Huang2014ComputingNP-complete}. Even when we limit ourselves to von Neumann measurements, the measurement operators for $J>1/2$ demand $2J \times (2J+1)$ parameters, which makes optimization impractical even for smaller $J$ \cite{Rossignoli2012MeasurementsSystems}.
In this study, we simplify the situation by considering only measurement sets made up of Euler rotation of the total angular momentum states. While this approach represents only a subset of the broadest possible measurements, our computation still offers an upper limit of discord. Despite its limitations, this method provides valuable insight into the behavior of discord, complementing the two-qubit result by minimizing the number of traced-out particles.

The Euler rotation has the following transformation rule on the angular momentum base:
\begin{align}
	&\mathcal{D(R(\phi,\theta,\chi))}  \ket{j,m} \\
	&= \sum_{m^\prime} \mathcal{D}_{m^\prime, m}^{(j)}(R(\phi,\theta,\chi)\ket{j,m^\prime} \label{eq:euler_rotation}\\
	&= \sum_{m^\prime} \exp{(i(m^\prime \alpha + m \gamma))} d_{m^\prime, m}^{(j)}(\beta) \ket{j,m^\prime}
\end{align}
where $d_{m^\prime, m}^{(j)}(\beta)$ is the Wigner d-matrices. With this relation, the measurement operators created by such operations are
\begin{align}
\pi_{m}^{j}(\phi, \theta)&=\mathcal{D(R(\phi,\theta,\chi))}\ket{J,M} \bra{J,M} \mathcal{D(R(\phi,\theta,\chi))} \nonumber \\
&= \sum_{n, \mu}e^{i\phi(\mu-n))} d_{n, m}^{j}(\theta)d_{\mu,m}^{j}(\theta)\ket{j,n}\bra{j,\mu}
\end{align}
It's important to note that $\chi$, appearing in \eqref{eq:euler_rotation}, doesn't affect the measurement operators. The operators essentially depend on two angular parameters, $\theta$ and $\phi$. We apply this measurement operator ($\pi^{ja}_m$) to subsystem A, comprising $2j_a$ particles, and compute the quantum discord based on \eqref{eq:discord}. 

\begin{figure}[h]
\centering
\includegraphics[width=0.9\columnwidth]{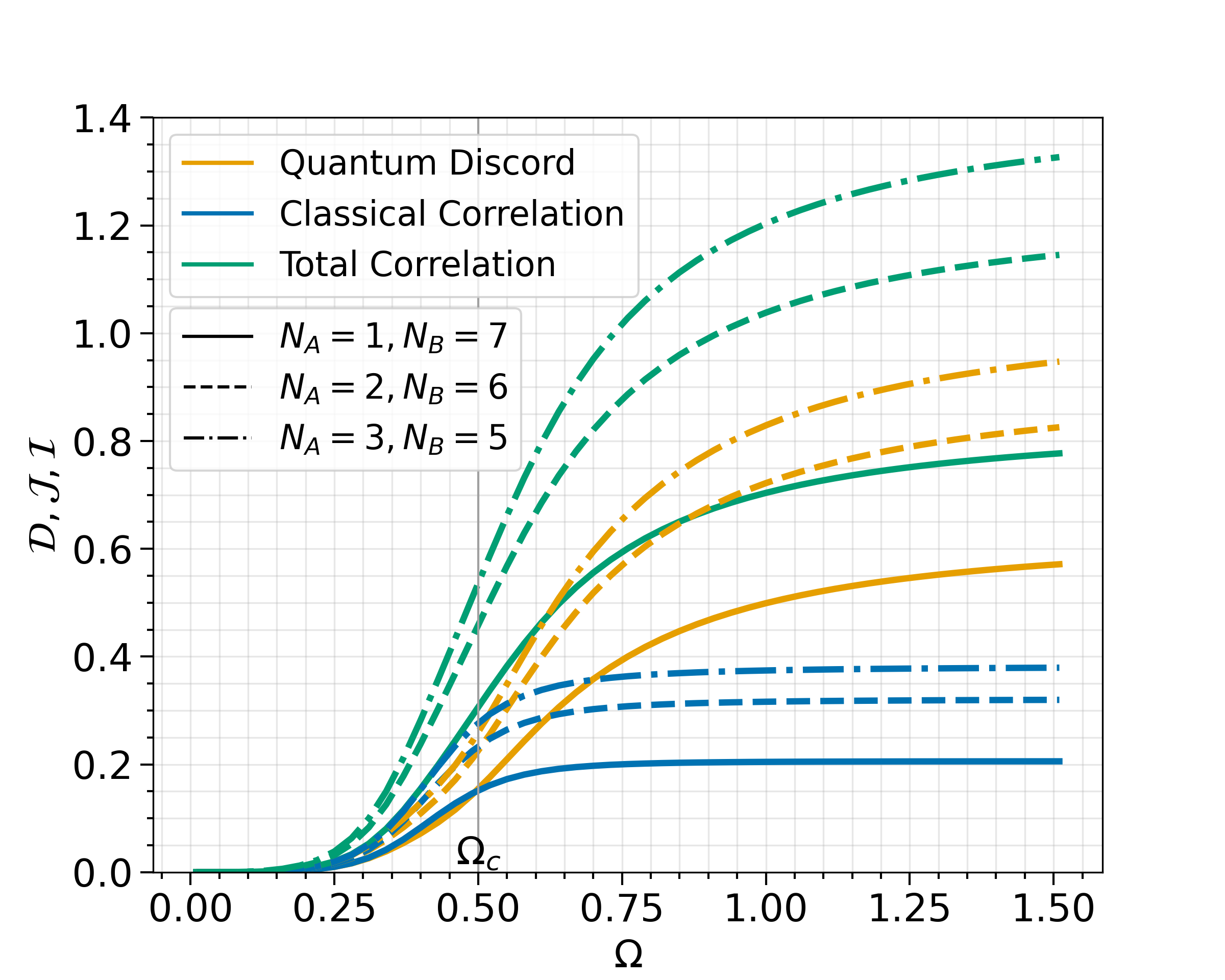}
\caption{\label{fig:dq_general_sum} (Color online) The upper limit of bipartite QD, the lower limit of classical correlation, and the total correlation of the system for $N = 8$ with different bipartition of the system. $N_A$ is the particle number of the system that was made measurement of, $N_B$ is the particle number of the other subsystem. (a)$N_A = 1$, (b) $N_A =2$, (c) $N_A = 3$. Unlike in \ref{fig:dq_qubit_sum}, no tracing is done on the density matrix. The overall behavior appears to be same as the 2-qubit case, suggesting QD is not sensitive to tracing operation in this model. }
\end{figure}

\fig{fig:dq_general_sum} shows the result plotted for the full density matrix. We observe that the bipartite discord behaves very similarly to the 2-qubit discord computed on the reduced density matrix. The tracing operation seems insignificant in the case of quantum disocrd in our case.  

\subsection{Global discord}
While the quantum discord we have utilized is applicable only for bipartite systems, it is certainly of interest to explore an extension of QD that includes multiple particles. One such extension is the Global Quantum Discord (GQD). This measure uses relative entropy, defined as $S(\rho| \sigma) = \operatorname{Tr}\left({\rho} \log _2 {\rho}-{\rho} \log _2 {\sigma}\right)$, and is formulated as follows:

\begin{align}
D^g\left({\rho}_{A_1 \cdots A_N}\right) 
&= \min _{\{{\Pi}_k\}} \left[ S\left({\rho}_{A_1 \cdots A_N} \| \Phi\left({\rho}_{A_1 \cdots A_N}\right)\right) \right. \nonumber \\
&\quad \left. - \sum_{j=1}^N S\left(\rho_{A_j} \| \Phi_j\left({\rho}_{A_j}\right)\right)\right]
\end{align}

where, $\Phi_j\left(\hat{\rho}_{A_j}\right)=\sum_{s} \hat{\Pi}_{A_j}^{s} \hat{\rho}_{A_j} \hat{\Pi}_{A_j}^{s}$ and $\Phi\left(\hat{\rho}_{A_1 \cdots A_N}\right)=$ $\sum_k \hat{\Pi}_k \hat{\rho}_{A_1 \cdots A_N} \hat{\Pi}_k$, with $\hat{\Pi}_k=\hat{\Pi}_{A_1}^{s_1} \otimes \cdots \otimes \hat{\Pi}_{A_N}^{s_N}$. Here, we consider each atom as a separate subsystem, designated $A_j$ for the $j$-th subsystem. For each measurement performed on subsystem $A_j$, we label it as $s_j$. Each subsystem consisting of a single atom includes two measurement operators, denoted as $s_j$ being either + or -. These operators are explicitly expressed as $\hat{\Pi}_{A_j}^{\pm}$ = $\ket{\pm}\bra{\pm}$, where the states $\ket{\pm}$ are defined in \eqref{eq:plus} \eqref{eq:minus}, replacing $\theta$ and $\phi$ with $\theta_j$ and $\phi_j$ respectively. The index string of $j$ is represented by $k$.

For $N=2$, GQD is equivalent to symmetric discord. Additionally, if the measurement operators are pre-selected, meaning that we choose the values of $\theta_j$ and $\phi_j$ instead of minimizing over them, GQD translates into Measurement-Induced Disturbance (MID) \cite{Luo2008UsingQuantum}. As these measures have detected and characterized phase transitions in certain systems where bipartite quantum discord (QD) does not \cite{Xu2014Measurement-inducedModel}, it is valuable to investigate GQD. In our study, we examined GQD for various values of $N$ and different configurations of $\theta_j$ and $\phi_j$.

\begin{figure}[h]
\centering
\includegraphics[width=0.9\columnwidth]{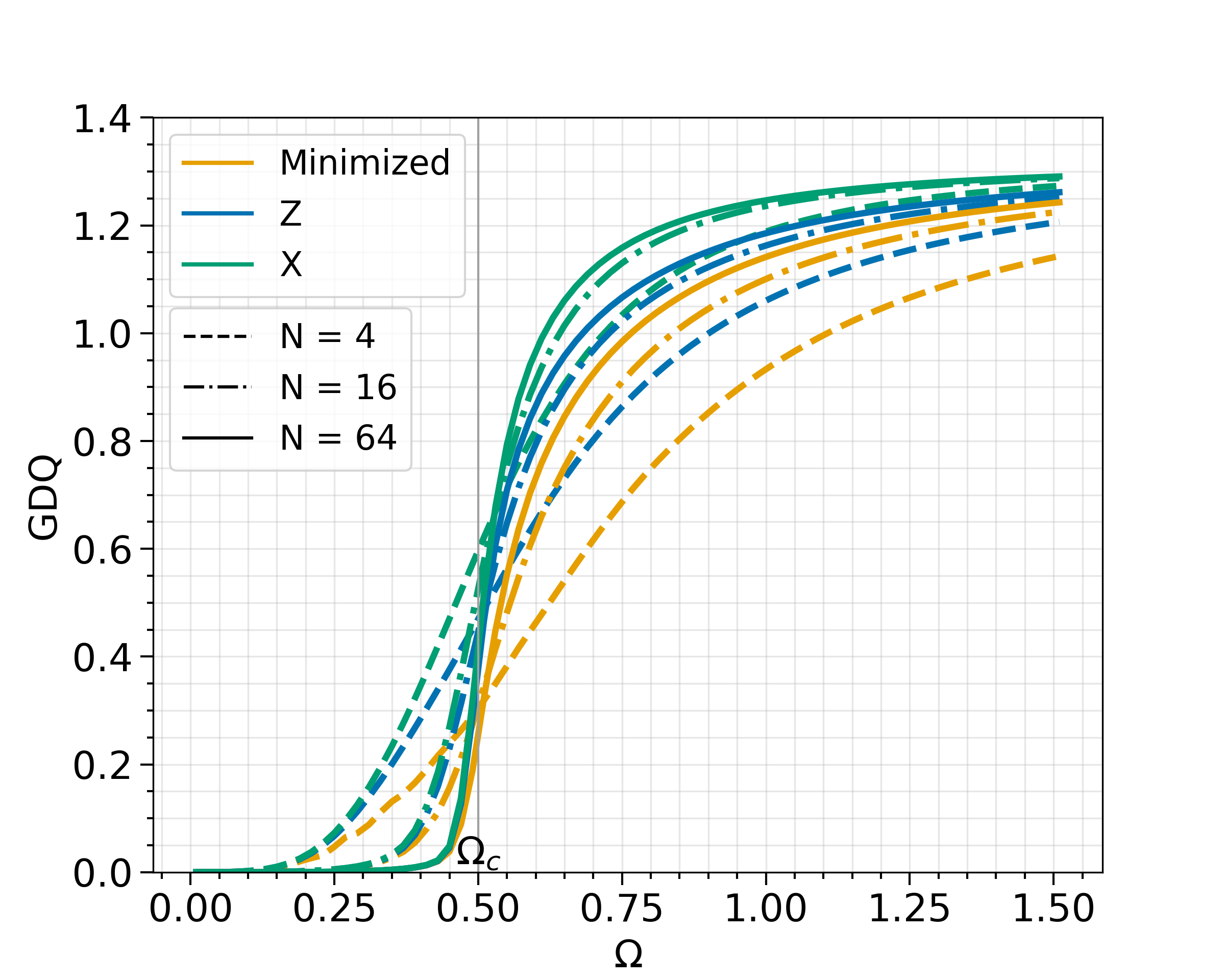}
\caption{\label{fig:global_dq} (Color online) 4-partite Global quantum discord plotted for the reduced density matrix of $N_R=4$ for different particle numbers of the total system. Each color represents the different measurement operator $\hat{\Pi}_k$ to compute $D_g$. Orange: optimized for minimum $D_g$, Blue: Spin measurement in Z direction (i.e. $\theta = \pi, \phi = 0$  in \eqref{eq:plus}, \eqref{eq:minus}), Green: Spin measurement in X direction ($\theta = \pi/2, \phi = 0$).    }
\end{figure}

The findings of our study, presented in Figure \fig{fig:global_dq}, demonstrate that Global Quantum Discord (GDQ) and Measurement-Induced Disturbance (MID) display characteristics similar to bipartite quantum discord, regardless of the number of particles in the chain. Contrary to the expected behavior, we observe that all variants of quantum discord exhibit a monotonic increase as the model's dissipation effect diminishes.

Typically, in pure quantum systems, quantum discord signals phase transitions, usually characterized by a smooth peak at the phase transition point or equivalently by the sign change in the first derivative of the discord. This behavior aligns intuitively, given that quantum discord serves as an inclusive measure of entanglement.

However, our results present a surprising contrast: quantum discord and quantum entanglement exhibit entirely different behaviors. This divergence could potentially be attributed to the system transitioning to a mixed state, where retaining quantum correlation as entanglement becomes more challenging, yet the correlation still persists. This observation prompts a more profound investigation into the complex dynamics of quantum correlations and entanglements.

\subsection{Multiparticle 'classical correlation' as a sign of phase transition}
While previous correlation measures based on mutual information showed similar monotonic increasing behavior, we report that multipartite mutual information following a measurement of the subsystem, provides a sharp signature of the phase transition. We present the results without a definitive interpretation.

For a 2-qubit quantum discord, the mutual information of the total system can be decomposed into two constituents: quantum discord ($\mathcal{D}$), quantifying the quantum correlation, and another component ($\mathcal{J}$), theoretically accounting for the classical correlation. We may interpret the later part as mutual information between subsystem a and b after the measurement is performed on subsystem a. This follows from the following equality: 
\begin{align}
    \mathcal{J(\varrho)}_{AB} &= S(\varrho_B) - \min_{\Pi^A_k \in \mathcal{M}^A} \sum_k p_k S(\varrho_{B/k})  \\
    &= \max_{\{{\Pi}_A\}}\mathcal{I}(\varrho_{\Pi_A}) 
    \label{eq:bigJ}
\end{align}
 where $\mathcal{I}$ is mutual information defined in \eqref{eq:total_correlation} and $\varrho_{{\Pi}_A} = \sum_k\Pi^A_k \varrho \Pi^A_K$ is the density matrix after a measurement is performed on the subsystem A. Already for a system with $N=2$, we observe that this quantity behaves distinctly from both $\mathcal{I}$ and $\mathcal{D}$ (See \fig{fig:dq_qubit_sum}).  Can we extend this to systems with larger $N$ to see how it scales with $N$? Let us apply the idea from multiparticle mutual information. In statistics, the mutual information has a multivariable extension:
\begin{align}
    I(A_1:A_2:...A_n) &= \sum_i H(A_i) - \sum_{i,j} H(A_i A_j) + ... \nonumber \\
    &+(-1)^n \sum_{i_1,i_2,...i_n}H(A_{i_1}A_{i_2}...A_{i_n})
\end{align}
For instance, $I(A:B:C) = H(A) + H(B) + H(C) - H(AB) - H(AC) + H(ABC)$.
 Let $\varrho^k$ denote the reduced density matrix where all but $k$ subsystems are traced out. By substituting $H$ with $S$ and $A_k$ with $\varrho^k$, we can naively define $\mathcal{I}$, the quantum version of mutual information for $N$ particles. Following the formulation of $\mathcal{J}$, we define the '$N$-partite classical correlation':
\begin{align}
    \mathcal{J}(\varrho)
    & \equiv \mathcal{I}(\varrho_{\Pi_A}) \\
    &= \sum_i S(\varrho^i_{\Pi_A}) - \sum_{i,j} S(\varrho^{i,j}_{\Pi_A}) + ... \nonumber \\ 
    &+(-1)^{n-1} \sum_{i_1,i_2,...i_n}S(\varrho^{i_1, i_2,...,i_n}_{\Pi_A})
\end{align}

\begin{figure}[h]
\centering
\includegraphics[width=1.0\linewidth]{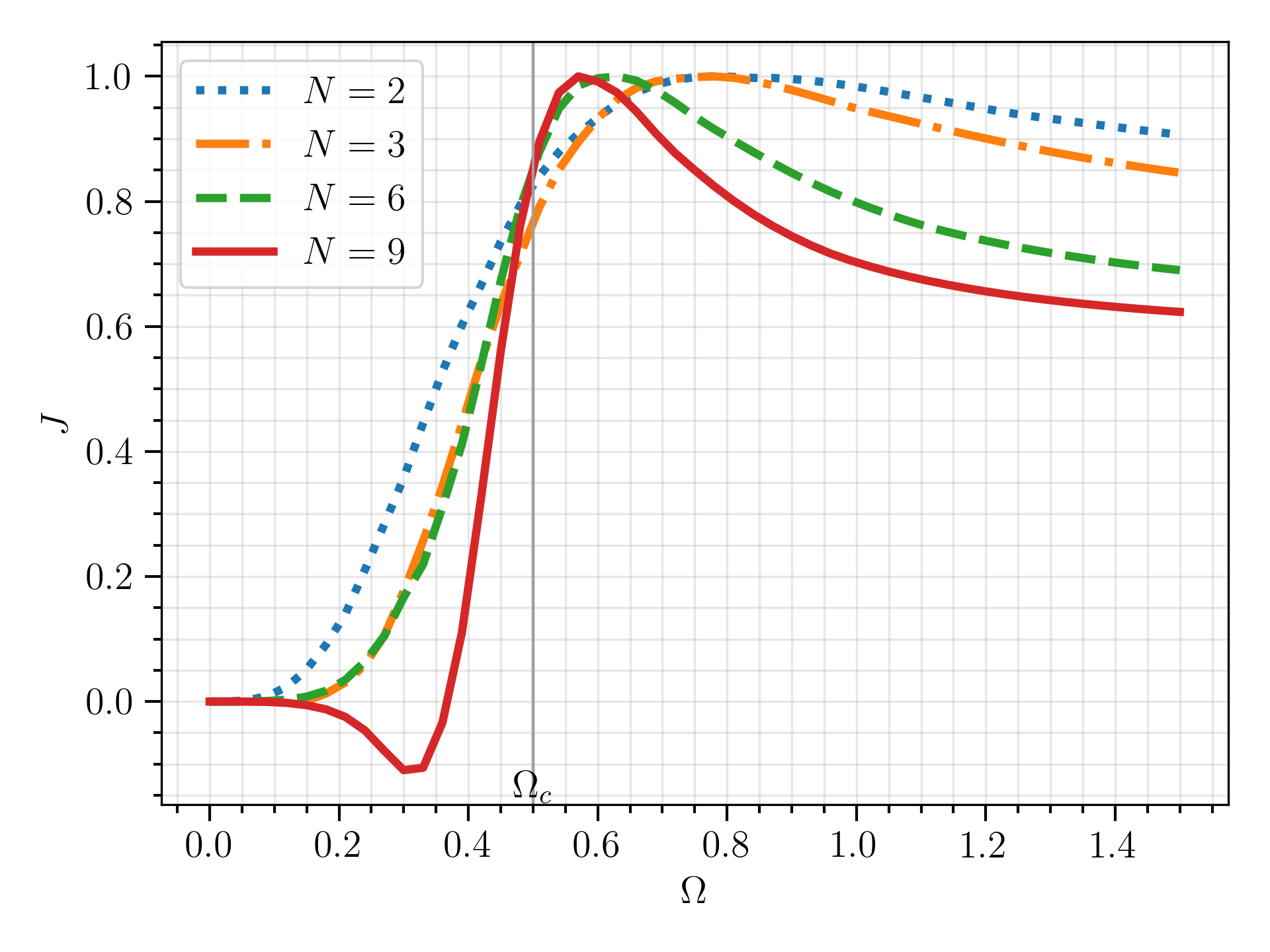}
\caption{\label{bigJ_1} (Color online) Rescaled multipartite "classical correlation" $J(\rho_{\pi A})$ of the system. The definition is given in \eqref{eq:bigJ}. $N$ represents the particle number of the entire system. The computation is done on full density matrix. (a) $N=2$, (b)$N=3$, (c)$N=6$, (d) $N=9$. For $N=2$, the plot is equivalent to the classical correlation defined in 2-qubit quantum discord \fig{fig:dq_qubit_sum}  }
\end{figure}

The '$N$-partite classical correlation' ($\mathcal{J}$) is depicted in \fig{bigJ_1}. 
Notably, $\mathcal{J}$ peaks at the critical transition point $\Omega_C$, contrasting the monotonic increase observed in the results from discord and mutual information. This finding suggests the emergence of classical-like correlations at the transition point, occurring simultaneously with the peaking of entanglement.

Interestingly, $\mathcal{J}$ yields finite values for correlated pure classical states (probabilistic mixture of product states), while it stays zero for many entangled states such as W state and Bell states. However, the precise interpretation of $\mathcal{J}$ remains elusive. Even within classical probability theory, the interpretation of multivariable mutual information is a topic of ongoing debate. Moreover, it has been demonstrated for $N=3$ that $\mathcal{J}$ fails to meet some fundamental properties of quantum measures \cite{Jin2019MonogamySystems}. Given this context, we present the results without asserting a definitive interpretation.

\section{Conclusion}
In this work, we conducted a detailed analysis of the properties of entanglement and quantum discord in the steady state of a driven-dissipative collective spin model. This model, dictated by the Lindbladian master equation, is recognized for displaying a dissipative phase transition, a phenomenon prompted by the interplay between its unitary and dissipative elements.

Despite the model's open nature, quantum entanglement peaks at the phase transition point, implying quantum fluctuations. On a mesoscopic scale, entanglement exists for both the pure phase ($\Omega < \Omega_C$) and the impure phase ($\Omega > \Omega_C$) near the phase transition point. However, at the thermodynamic limit, the NPT entanglement of the latter case disappears. The presence of entanglement in the impure phase on a mesoscopic scale is particularly significant, given that the proposed experimental setup of the model assumes this scale. Due to the symmetric nature of this model, all entanglement is classified as genuinely multiparticle entanglement (GME). Our study further reveals that the density matrix becomes separable when $\Omega \to \infty$ and $\Omega \to 0$, showing that entanglement decays and asymptotically approaches zero as $\Omega$ diverges from the phase transition point.

Interestingly, we observe a contrasting behavior in the case of quantum discord and its variants, all of which increase monotonically with $\Omega$. This discrepancy between the behaviors of quantum discord and entanglement is intriguing, as quantum discord represents a comprehensive measure of non-classical correlations, including entanglement itself.

Based on these observations, we interpret the behavior of quantum correlations as follows: As the Hamiltonian term intensifies, so does the quantum correlation. In the dissipation-dominant regime of the pure phase, this increase in quantum correlation manifests as entanglement. However, as the system transitions into the impure phase, the entanglement component of the quantum correlation rapidly decreases. As the Hamiltonian term further strengthens, this entanglement eventually vanishes, leaving quantum discord to account for most of the quantum correlation.

We also draw attention to a quantity presumably associated with classical correlation, which peaks and shows signs of discontinuity at the phase transition point. This suggests that the phase transition encompasses not only quantum fluctuations, primarily evident as entanglement, but also classical-like fluctuations driven by the dissipation process.

While the primary aim of this paper was to elucidate the relationship between correlation measures and the dissipative phase transition, it has additionally highlighted that, experimentally, this model serves as an intriguing platform for various types of atom-atom correlations. In practical experiments, our findings suggest that by modulating a single parameter - essentially the Rabi frequency - one can access a spectrum of entanglement types, as well as the quantum discord of the system.

\section*{ACKNOWLEDGMENTS}
We want to thank Longbao Xiao, Joseph Wheeler and Kyosuke Higuchi for discussions. This work was supported by the NSF through the CUA PFC and the AFOSR. Q.W. also acknowledges financial support from Japan Student Services Organization (JASSO).

\bibliography{references}

\appendix

\section{Proof $\rho_{ss}(\Omega \to 0)$ and $\rho_{ss}(\Omega \to \infty)$ are separable} \label{app:proof_max_mix}
We prove the following state is separable and therefore not entangled. \begin{equation} \lim_{\Omega \to \infty} \rho_{ex}(\Omega) = \frac{1}{N}\sum_{m = -J}^{J} \ket{J, m} \bra{J, m}. \label{eq:infsol} \end{equation} 
Such states generally take the form \begin{equation} \rho = \sum _ { n = 0 } ^ { N } \chi _ { n } | D _ { N , n } \rangle \left\langle D _ { N , n } |\right.. \label{eq:mixDicke} \end{equation}, 
Here, the basis $ | D _ { N , n } \rangle$ comprises the unnormalized Dicke states, 
which are related to our angular momentum basis as \begin{equation} \ket{J,m} = \binom{2J}{J-M} \ket{D_{2J,J-m}}. \label{eq:basistrans} \end{equation}
where $\binom{n}{m} = \frac{n!}{(n-m)!m!}$ is the binomial coefficient.

A necessary and sufficient condition for the separability of such diagonal symmetric states is that the following Hankel matrices constructed using the matrix component $\Xi$ in \eqref{eq:mixDicke} are positive semidefinite \cite{Yu2016}:
\begin{align} 
H _ { 0 } & : = \left( \begin{array} { c c c } 
\chi _ { 0 } & \cdots & \chi _ { m _ { 0 } } \\
\cdots & \cdots & \cdots \\
\chi _ { m _ { 0 } } & \cdots & \chi _ { 2 m _ { 0 } } 
\end{array} \right), \\ 
H _ { 1 } & : = \left( \begin{array} { c c c } 
\chi _ { 1 } & \cdots & \chi _ { m _ { 1 } } \\
\cdots & \cdots & \cdots \\
\chi _ { m _ { 1 } } & \cdots & \chi _ { 2 m _ { 1 } - 1 } 
\end{array} \right). 
\end{align}

The components of the Hankel matrices corresponding to $\rho_{ss}(\Omega \to \infty)$ can be obtained from Eqs.\eqref{eq:infsol} and \eqref{eq:basistrans} as
\begin{equation}
\chi_k = \frac{1}{2J+1}\binom{2J}{k}^{-1}
\end{equation}
Therefore the $(i,j)$ components of Hankel matrices are written as,
\begin{align}
    (H_0)_{i,j} &= \chi_{i+j} = \frac{1}{2J+1} \binom{2J}{i+j}^{-1} \\
    (H_1)_{i,j} &= \chi_{i+j+1} = \frac{1}{2J+1} \binom{2J}{i+j+1}^{-1}
\end{align}
The range of indices are $0 \leq i,j \leq \lfloor J \rfloor$ for $H_0$ and $0 \leq i,j \leq \lfloor J - 1/2 \rfloor$ for $H_1$, obtained from $m_0$ and $m_1$. The $\lfloor \quad \rfloor$ represents the integer floor function. 

We show that these matrices are indeed positive semidefinite by explicitly showing these two matrices are Gram matrices. 
Writing binomial coefficient explicitly using factorials, we can write the components of the matrices as follows, 
\begin{equation}
    (\Tilde{H}_\delta)_{i,j} = \frac{(i+j+\delta)!(2J-i-j)! }{(2J+1)!}
\end{equation}
where $\delta$ is either 0 or 1.
Let's define set of functions $P_k(x)$ labeled by an integer $k$ 
\begin{equation}
    P_k(x) = x^{\frac{\delta}{2}}\left( \frac{x}{1-x} \right)^k
\end{equation}
We define the inner product of these functions $\langle P_i(x) P_j(x) \rangle$ using the weight function $w(x) = (1-x)^{2J}$. It evaluates to
\begin{align}
\langle P_i(x) P_j(x) \rangle
&= \int_0^1 P_i(x) P_j(x) w(x) dx \\
&= \int_0^1 x^{i+j+\delta} (1-x)^{2J - i -j} dx \\
&= B(i+j+1, 2J-i-j+1) \\
&= \frac{(i+j+\delta)!(2n-i-j)!}{(2J+1)!} \\
&= (\Tilde{H}_\delta)_{i,j}
\end{align}
where we used the definition of beta functions. Indeed, $H_0$ and $H_1$ are gram matrices for all $J > 1/2$, and therefore positive semidefinite. This concludes the proof that density matrix $\rho_{ss}(\Omega \to \infty$) is separable.

\end{document}